\documentclass[aps,preprint]{revtex4}

\usepackage{graphicx}

\begin{document}

\title{The origin of bursts and heavy tails in human dynamics}

\author{Albert-L\'aszl\'o Barab\'asi}

\affiliation{Center for Complex Networks Research and Department
of Physics, University of Notre Dame, IN 46556, USA}

\date{\today}

\vskip 4truecm

\begin{abstract}
{\bf   The dynamics of many social, technological and economic
phenomena are driven by individual human actions, turning the
quantitative understanding of human behavior into a central
question of modern science. Current models of human dynamics, used
from risk assessment to communications, assume that human actions
are randomly distributed in time and thus well approximated by
Poisson processes \cite{Haight67,call-center,inventory}. In
contrast, there is increasing evidence that the timing of many
human activities, ranging from communication to entertainment and
work patterns, follow non-Poisson statistics, characterized by
bursts of rapidly occurring events separated by long periods of
inactivity
\cite{phone-design,Instant,supercomputers,ftp,economic2}. Here we
show that the bursty nature of human behavior is a
 consequence of a decision based queuing process
\cite{Cobham,queue-cohen}: when individuals execute tasks based on
some perceived priority, the timing of the tasks will be heavy
tailed, most tasks being rapidly executed, while a few experience
very long waiting times. In contrast, priority blind execution is
well approximated by uniform interevent statistics. These findings
have important implications from resource management to service
allocation in both communications and retail.}
\end{abstract}

%\pacs{}

\maketitle

Humans participate on a daily basis in a large number of distinct activities, ranging from electronic
communication, such as sending emails or making phone calls, to browsing the web, initiating financial
transactions, or engaging in entertainment and sports.
 Given the number of factors that determine the timing of
each action, ranging from work and sleep patterns to resource availability,
it appears impossible to seek
regularities in human dynamics, apart from the obvious daily and seasonal periodicities. Therefore, in contrast
with the accurate predictive tools common in physical sciences, forecasting human and social patterns remains a
difficult and often elusive goal.

\begin{figure}[!b]
\centerline{\includegraphics[width=14.0cm]{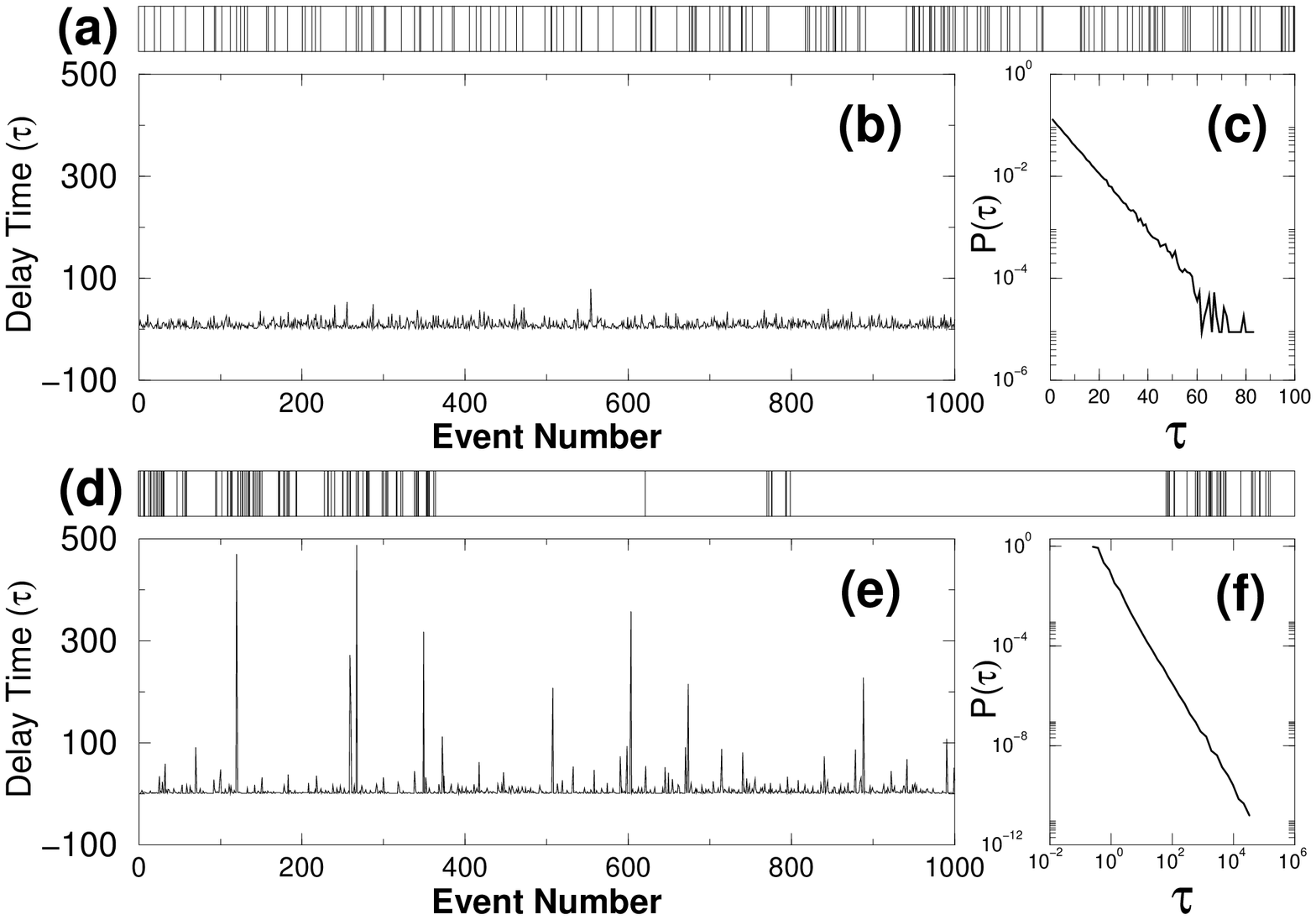}}
\caption{\small {The difference between the activity patterns
predicted by a Poisson process (top) and the heavy tailed
distributions observed in human dynamics (bottom). {\bf (a)}
Succession of events predicted by a Poisson process, which assumes
that in any moment an event takes place with probability $q$. The
horizontal axis denotes time, each vertical line corresponding to
an individual event. Note that the interevent times are comparable
to each other, long delays being virtually absent. {\bf (b)} The
absence of long delays is visible on the plot showing the delay
times $\tau$ for 1,000 consecutive events, the size of each
vertical line corresponding to the
 gaps seen in (a).  {\bf (c)} The probability of finding exactly $n$ events within a fixed time
interval is $ {\cal P}(n;q)= e^{-q t} (q t)^n/n!$, which predicts
that for a Poisson process the inter-event time distribution
follows $P(\tau)=q e^{-q\tau}$, shown on a log-linear plot in (c)
for the events displayed in (a, b). {\bf (d)} The succession of
events for a heavy tailed distribution.
 {\bf (e)} The waiting time $\tau$ of 1,000 consecutive events, where the
mean event time was chosen to coincide with the mean event time of
the Poisson process shown in (a-c). Note the large spikes in the
plot, corresponding to very long delay times.  (b) and (e) have
the same vertical scale, allowing to compare the regularity of a
Poisson process  with the bursty nature of the heavy tailed
process.  {\bf (f)} Delay time distribution $P(\tau) \simeq
\tau^{-2}$ for the heavy tailed process shown in (d,e), appearing
as a straight line with slope -2 on a log-log plot. The signal
shown in (d-f) was generated using $\gamma=1$ in the stochastic
priority list model discussed in the Supplementary Information.}}
\label{fig1}
\end{figure}

Current models of human activity are based on Poisson processes,
and assume that in a $dt$ time interval an individual (agent)
engages in a specific action with probability $qdt$, where $q$ is
the overall frequency of the monitored activity. This model
predicts that the time interval between  two consecutive actions
by the same individual, called the waiting or inter-event time,
follows an exponential distribution (Fig. \ref{fig1}, a-c)
\cite{Haight67}. Poisson processes are widely used to quantify the
consequences of human actions, such as modelling traffic flow
patterns or accident frequencies \cite{Haight67}, and are
commercially used in call center staffing \cite{call-center},
inventory control \cite{inventory}, or to estimate the number of
congestion caused blocked calls in mobile communications
\cite{phone-design}. Yet, an increasing number of recent
measurements indicate that the timing of many human actions
systematically deviate from the Poisson prediction, the waiting or
inter-event times being better approximated by a heavy tailed or
Pareto distribution (Fig. \ref{fig1}, d-f). The differences
between Poisson and heavy tailed behavior is striking: a Poisson
distribution decreases exponentially, forcing the consecutive
events to follow each other at relatively regular time intervals
and forbidding very long waiting times. In contrast, the slowly
decaying heavy tailed processes allow for very long periods of
inactivity that separate bursts of intensive activity (Fig. 1).

To provide direct evidence for non-Poisson activity patterns in
individual human behavior, we study the communication between
several thousand  email users based on a dataset capturing the
sender, recipient, time and size of each email
\cite{eckmann,ebel}. As Figure 2a shows, the distribution of the
time differences between consecutive emails sent by a selected
user is best approximated with  $P(\tau)\sim \tau^{-\alpha}$,
where $\alpha \simeq 1$, indicating that an individual's email
pattern has a bursty non-Poisson character: during a single
session a user sends out several emails in a quick succession,
followed by  long periods of no email activity. This behavior is
not limited to email communications. Measurements capturing the
distribution of the time differences between consecutive instant
messages sent by individuals during online chats \cite{Instant}
show a similar pattern. Professional tasks, such as the timing of
job submissions on a supercomputer \cite{supercomputers},
directory listings and file transfers (FTP requests) initiated by
individual users \cite{ftp}, or the timing of printing jobs
submitted by users \cite{Maya} were also reported to display
non-Poisson features. Similar patterns emerge in the time interval
distribution between individual trades in currency futures
\cite{economic2}. Finally, heavy tailed distributions characterize
entertainment related events, such as the time intervals between
consecutive online games played by the same user \cite{games}.

 The fact that a wide range of human activity patterns
follow non-Poisson statistics suggests that the observed bursty
character reflects some fundamental and potentially generic
feature of human dynamics. Yet, the mechanism responsible for
these striking non-random features remain unknown. Here we show
that the bursty nature of human dynamics is a consequence of a
queuing process driven by human decision making: whenever an
individual is presented with multiple tasks and chooses among them
based on some perceived priority parameter, the waiting time of
the various tasks will be Pareto distributed. In contrast,
first-come-first-serve and random task execution, common in most
service oriented or computer driven environments, lead to a
uniform Poisson like dynamics.

Most human initiated events require an individual to weigh and
prioritize different activities. For example, at the end of each
activity an individual needs to decide what to do next: send an
email, do some shopping, or place a phone call, allocating time
and resources for the chosen activity. Consider an agent operating
with a priority list of $L$ tasks. After a task is executed, it is
removed from the list, offering the opportunity to add another
task. The agent assigns to each task a priority parameter $x$,
which allows it to compare the urgency of the different tasks on
the list.  The question is, how long will a given task have to
wait before it is executed. The answer depends on the method the
agent uses to choose the task to be  executed next.  In this
respect three selection protocols \cite{queue-cohen} are
particularly relevant for human dynamics:

(i) The simplest is the first-in-first-out protocol, executing the
tasks in the order they were added to the list. This is common in
service oriented processes, like the first-come-first-serve
execution of orders in a restaurant or getting help from directory
assistance and consumer support. The time period an item stays on
the list before execution is determined by the cumulative time
required to perform all tasks added to the list before it. If the
time necessary to perform the individual tasks are chosen from a
bounded distribution ({\it i.e.} the second moment of the
distribution is finite), then the waiting time distribution will
develop an exponential tail, indicating that most tasks experience
approximately the same waiting time.

(ii) The second possibility is to execute the tasks in a random
order, irrespective of their priority or time spent on the list.
This is common, for example, in educational settings, when
students are called on randomly, and in some packet routing
protocols. The waiting time distribution of the individual tasks
(i.e. the time between two calls on the same student) in this case
is also exponential.

(iii)  In most human initiated activities task selection is not
random, but the individual executes the highest priority item on
its list. The resulting execution dynamics is quite different from
(i) and (ii): high priority tasks will be executed soon after
their addition to the list, while low priority items will have to
wait until all higher priority tasks are cleared, forcing them to
stay on the list for considerable time intervals. In the following
we show that this selection mechanism, practiced by humans on a
daily basis, is the likely source of the fat tails observed in
human initiated processes.

We assume that an individual has a priority list with $L$ tasks,
each  task being assigned a priority parameter $x_i, ~i=1,..., L$,
chosen  from a $\rho(x)$ distribution. At each time step the agent
selects the highest priority task from the list and executes it,
removing it from the list. At that moment a new task is added to
the list, its priority $x_i$ being again chosen from $\rho(x)$.
This simple model ignores the possibility that the agent
occasionally selects a low priority item for execution before all
higher priority items are done, common, for example, for tasks
with deadlines. This can be incorporated by assuming that the
agent executes the highest priority item with probability $p$, and
with probability $1-p$ executes a randomly selected task,
independent of its priority. Thus the $p \to 1$ limit of the model
describes the deterministic (iii) protocol, when always the
highest priority task is chosen for execution, while $p \to 0$
corresponds to the random choice protocol discussed in (ii).

To establish that this priority list model can account for the
observed fat tailed interevent time distribution, we first studied
its dynamics numerically with priorities chosen from a uniform
distribution $x_i \in [0,1]$. Computer simulations show that in
the $p \to 1$ limit the probability that a task spends $\tau$ time
on the list has a power law tail with exponent $\alpha=1$ (Fig
3a), in agreement with the exponent obtained in email
communications (Fig 2a). In the $p\to 0$ limit $P(\tau)$ follows
an exponential distribution (Fig. 3b), as expected for the case
(ii). As the typical length of the priority list differs from
individual to individual, it is particularly important for the
tail of $P(\tau)$ to be independent of $L$. Numerical simulations
indicate that this is indeed the case: changes in $L$ do not
affect the scaling of $P(\tau)$. The fact that the scaling holds
for $L=2$ indicates that it is not necessary to have a long
priority list: as long as individuals balance at least two tasks,
a bursty heavy tailed interevent dynamics will emerge.

 To determine the tail of  $P(\tau)$ analytically we consider a
 stochastic version of the model in which the probability to choose a
task with priority $x$ for execution in a unit time is  $\Pi(x)
\sim x^\gamma$, where $\gamma$ is a parameter that allows us to
interpolate between the random choice limit (ii) ($\gamma=0$,
$p=0$) and the deterministic case, when always the highest
priority item is chosen for execution (iii) ($\gamma=\infty$,
$p=1$). Note that this parameterization captures the scaling of
the model only in the $p \to 0$ and $p\to 1$ limits, but not for
intermediate $p$ values, thus it is chosen only for mathematical
convenience. The probability that a task with priority $x$ waits a
time interval $t$ before execution is ${f
}(x,t)=(1-\Pi(x))^{t-1}\Pi(x)$. The average waiting time of a task
with priority $x$ is obtained by averaging over $t$ weighted with
 ${f} (x,t)$, providing
\begin{equation}
\tau(x)=\sum_{t=1}^\infty t f(x,t) =\frac{1}{\Pi(x)}\sim \frac{1}
{x^\gamma}, \label{tx}
\end{equation}
i.e. the higher an item's priority, the shorter is the average
time it waits before execution. To calculate  $P(\tau)$  we use
the fact that the priorities are chosen from the $\rho(x)$
distribution, i.e. $\rho(x)dx=P(\tau) d\tau$, which gives
\begin{equation}
P(\tau) \sim \frac{\rho(\tau^{-1/\gamma})}{\tau^{1+1/\gamma}}.
\label{Pt}
\end{equation}
In the $\gamma\to\infty$ limit, which converges to the strictly
priority based deterministic choice ($p=1$) in the model, Eq.
(\ref{Pt}) predicts  $P(\tau) \sim \tau^{-1}$, in agreement with
the numerical results (Fig 3a), as well as the empirical data on
the email interarrival times (Fig 2a). In the $\gamma=0$ ($p=0$)
limit $\tau(x)$ is independent of $x$, thus $P(\tau)$ converges to
an exponential distribution, as shown in Fig. 3b (see
Supplementary Information).

The apparent dependence of $P(\tau)$ on the $\rho(x)$ distribution
from which the agent chooses the priorities may appear to
represent a potential problem, as assigning priorities is a
subjective process, each individual being characterized by its own
$\rho(x)$ distribution. According to Eq. (\ref{Pt}), however, in
the $\gamma \to \infty$ limit $P(\tau)$ is independent of
$\rho(x)$. Indeed, in the deterministic limit the uniform
$\rho(x)$ can be transformed into an arbitrary $\rho'(x)$ with a
parameter change, without altering the order in which the tasks
are executed \cite{queue-cohen}. This insensitivity of the tail to
$\rho(x)$ explains why, despite the diversity of human actions,
encompassing both professional and personal priorities, most
decision driven processes develop a heavy tail.

To obtain empirical evidence for the validity of the proposed
queuing mechanism we consider the email activity pattern of an
individual \cite{eckmann,ebel}. Once in front of the computer, an
individual will reply immediately to a high priority message,
while placing the less urgent or the more difficult ones on its
priority list to compete with other non-email activities.  We
propose, therefore, that the observed interevent time distribution
is in fact rooted in the uneven waiting times experienced by
different tasks. To test this hypothesis we need to determine
directly the waiting time for each task. In the email dataset we
have the time, sender and recipient of each email transmitted over
several months by each user, thus we can determine the time it
takes for a user to reply to a received message \cite{eckmann}. As
Fig. 2b shows, we find that the waiting time distribution
$P(\tau_w)$ for the user whose $P(\tau)$ is shown in Fig. 2a is
best approximated by $P(\tau_w) \sim \tau_w^{-\alpha_w}$ with
exponent $\alpha_w=1$, supporting our hypothesis that the heavy
tailed waiting time distribution drives the observed bursty email
activity patterns.

As in the $p \rightarrow1$ limit of the model the priority list is
dominated by low priority tasks, new tasks will often be executed
immediately. This results in a peak at $P(\tau=1)$(see Fig. 3 in
the Supplementary Information), which, while in some cases may
represent a model artifact, in the email context is not
unrealistic: most emails are either deleted right away (which is
one kind of task execution), or are immediately replied to. Only
the more difficult or time consuming tasks will queue on the
priority list. The email dataset does not allow us to resolve this
peak, however: a message to which the user replies right away will
appear to have some waiting time, given the delay between the
arrival of the message and the time the user has a chance to read
it.

While we illustrated the queuing process on emails, in general the
model is better suited to capture the competition between
different kinds of activities an individual is engaged in, i.e.
the switching between various work, entertainment and
communication events. Indeed, most datasets displaying heavy
tailed interevent times in a specific activity reflect the outcome
of the competition between tasks of different nature. For example,
the starting of an online gaming session implies that all higher
priority work and entertainment related activities have been
already executed.

Detailed models of human activity require us to consider the
impact of a number of additional mechanisms on the queuing
process. First, in the priority list model we assumed that the
time necessary to execute a task (service time) is the same for
all tasks. The size distribution of emails is heavy tailed
\cite{Crovella,Mitzenmacher}, however, thus the waiting time
distribution could be driven entirely by the time it takes to read
an email, i.e. the message size. Yet, as Fig. 2c shows, we fail to
find a correlation between the size of the email received by a
user and the time the user takes to reply to it. While a detailed
analysis should also consider the role of attachments, Fig. 2c
suggests that the priority of a response is more important than
the message size. Furthermore, the priorities assigned to tasks
are often driven by optimization processes, as agents aim to
maximize profits or minimize the overall time spent on some
activity.

A natural extension of the model is to assume that tasks arrive at
a rate $\lambda$ and are executed at a rate $\mu$, allowing the
length of the priority list $L$ to change in time. In this case
the model maps into Cobham's priority queue model \cite{Cobham},
which has a power law distributed waiting time with $\alpha=3/2$
only when $\lambda=\mu$ (see Supplementary Information). Thus to
account for the power law waiting times
 the model requires an additional mechanism that  guarantees
 $\lambda=\mu$
 (which, as Fig. 3d indicates, is not satisfied for most email users).
In contrast, in the proposed priority list model we assumed that
for humans the length of the priority list remains relatively
unchanged (i.e. $L$ is constant). To understand the origin of this
assumption we must realize that for $\lambda=\mu$ the length of
the priority list fluctuates widely and can occasionally grow very
long. While keeping track of a long priority list is not a problem
for a computer, it is well established that the immediate memory
of humans has finite capacity \cite{Miller}. In other words, the
number of priorities we can easily remember, and therefore the
length of the priority list, is bounded, motivating our choice of
a finite $L$.

While other generalizations are possible and often required, our
main finding is that the observed fat tailed activity
distributions can be explained by a simple hypothesis: humans
execute their tasks based on some perceived priority, setting up
queues that generate very uneven waiting time distributions for
different tasks. In this respect the proposed priority list model
represents only a minimal framework that allows us to demonstrate
the potential origin of the heavy tailed activity patterns, and
offers room for further extensions to capture more complex human
behavior. As the exponent of the tail could depend on the details
of the prioritizing process, future work may allow the empirical
data to discriminate between different queuing hypotheses. A
mapping into punctuated equilibrium models (see Supplementary
Information \cite{Sneppen,Jensen}), with the mathematical
framework of queuing theory could help the systematic
classification of the various temporal patterns generated by human
behavior.

There is overwhelming evidence that Internet traffic is
characterized by heavy-tailed statistics \cite{Park}, rooted in
the Pareto size distribution of the transmitted files
\cite{Crovella,Mitzenmacher}. As we have shown above (Fig 2c), we
find that a user's email activity  does not correlate with the
email size. Similarly, the timing of online games \cite{games} or
sending an instant message \cite{Instant} cannot be driven by file
sizes either. This suggests that Internet traffic is in fact
driven by two separate processes: The heavy tailed size
distribution of the files sent by the users and the human decision
driven timing of various Internet mediated activities individuals
engage in. In some environments this second mechanism, whose
origin is addressed in this paper, can be just as important as the
much investigated first one. Given the differences in routing
performance under Poisson and Pareto arrival time distributions
\cite{Park,ERLANG-X,Leighton}, a queuing based model of
human-driven arrival times could contribute to a better
understanding of Internet traffic as well.

Uncovering the mechanisms governing the timing of various human
activities has significant  scientific and commercial potential.
First, models of human behavior are indispensable for large-scale
models of social organization, ranging from detailed urban models
\cite{toro-paper,manrubia}, to modeling the spread of epidemics
and viruses, the development of panic \cite{helbing} or capturing
financial market behavior \cite{caldarelli}. Understanding the
origin of the non-Poisson nature of human dynamics could
fundamentally alter the dynamical conclusions these models offer.
Second, models of human behavior are crucial for better resource
allocation and pricing plans for phone companies, to improve
inventory and service allocation in both online and
brick-and-mortal retail, and potentially to understand the bursts
of ideas and memes  emerging in communication and publication
patterns \cite{Kleinberg}. Finally, heavy tails have been observed
in the foraging patterns of birds as well \cite{viswanathan},
raising the intriguing possibility that animals also utilize some
evolution-encoded priority based queuing mechanisms to decide
between competing tasks, like caring for offsprings, gathering
food, or fighting off predators.

\pagebreak

{\bf Supplementary Information} accompanies the paper on {\bf
www.nature.com/nature}

{\bf Acknowledgements} I have greatly benefited from discussions
with Alexei Vazquez on the mathematical aspects of the model. I
also thank Luis Amaral, Zolt\'an Dezs\"o, Plamen Ivanov, Janet
Kelley, J\'anos Kert\'esz, Adilson Motter, Maya Paczuski, Kim
Sneppen, Tam\'as Vicsek, Ward Whitt, and Eduardo Zambrano for
useful discussions and comments on the manuscript, Jean-Pierre
Eckmann for providing the email database and  Suzanne Aleva for
assisting me with the manuscript preparation. This research was
supported by NSF DMR04-26737 and NSF DMS04-41089.

{\bf Competing interests statement} The author declares that he
has no competing financial interests.

{\bf  Correspondence} and requests for materials  should be
addressed to A.-L. B. (alb@nd.edu).

\pagebreak

\begin{figure}[!b]
\centerline{\includegraphics[width=13.0cm]{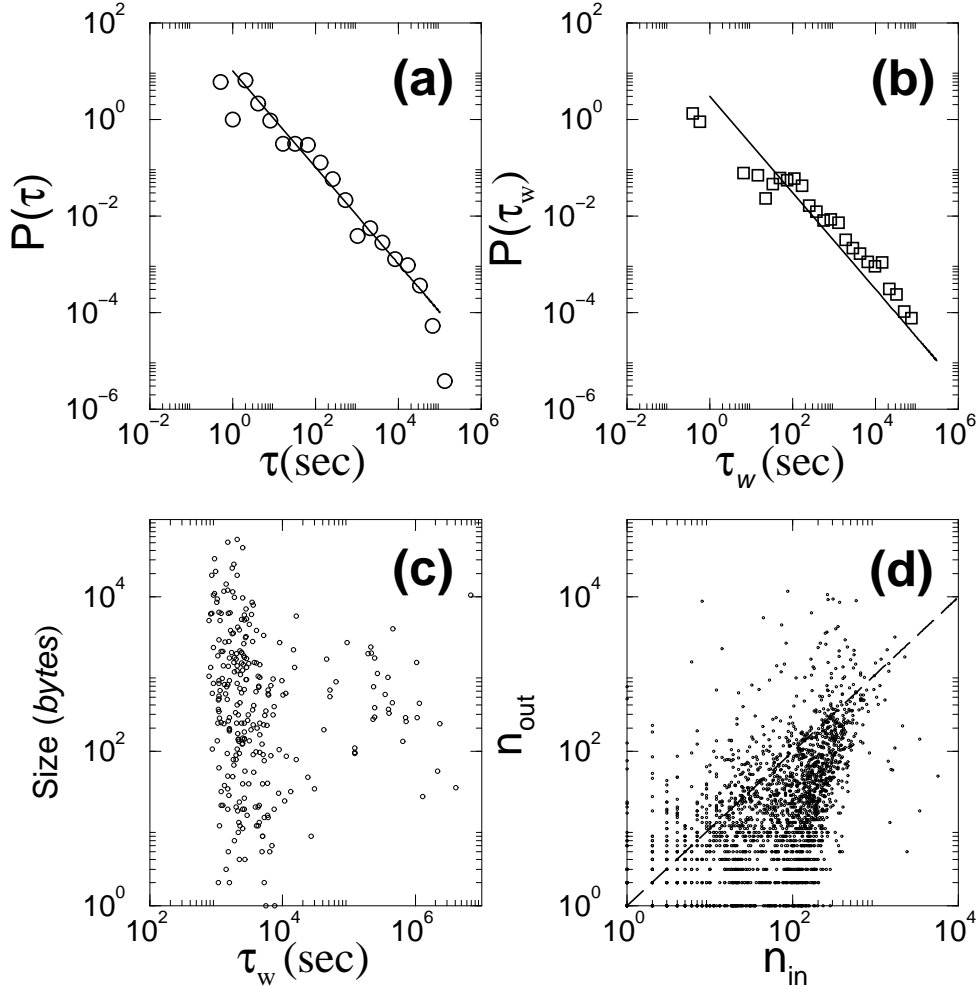}}
\caption{\small{Heavy tailed activity patterns in email
communications. {\bf (a)} The  distribution of the time intervals
between consecutive emails sent by a single user over a three
month time interval, indicating that $P(\tau) \sim \tau^{-1}$ (the
solid line in the log-log plot has slope -1). While the exponent
differs slightly from user to user, it is typically centered
around $\alpha=1$. {\bf (b)} The distribution of the time taken by
the user to reply to a received message.  To determine $\tau_w$ we
recorded the time the user received an email from a specific user,
and the time it sent a response to it, the difference between the
two providing $\tau_w$. For consistency the figure shows the data
for the  user whose interevent time distribution is shown in (a).
The solid line in the log-log plot has slope -1. {\bf (c)}  A
scatter plot showing the waiting time ($\tau_w$) and the size for
each email responded to by the user discussed in (a,b), indicating
that the file size and the response time do not correlate. {\bf
(d)} Scatter plot showing the number of emails received and sent
by 3,188 users during a three month interval. Each point
corresponds to a different user, indicating that there are
significant differences between the number of received and
responded emails. The dashed line corresponds to $n_{in}=n_{out}$,
capturing the case when the classical queuing models also predict
a power law waiting time distribution (see Supplementary
Information), albeit with exponent $\alpha=3/2$.}} \label{fig2}
\end{figure}

\begin{figure}[!b]
\centerline{\includegraphics[width=13.0cm]{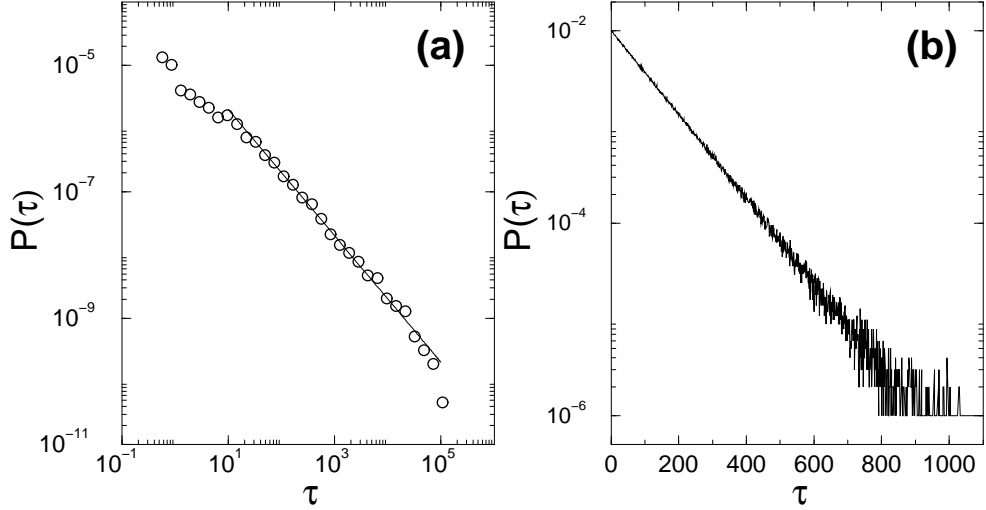}}
\caption{\small {The waiting time distribution predicted by the
investigated queuing model. The priorities were chosen from a
uniform distribution $x_i \in [0,1]$, and we monitored a priority
list of length $L=100$ over $T=10^6$ time steps. {\bf (a)} Log-log
plot of the tail of probability $P(\tau)$ that a task spends
$\tau$ time on the list obtained for $p=0.99999$, corresponding to
the deterministic limit of the model. The continuous line on the
log-log plot correspond to the scaling predicted by (\ref{Pt}),
having slope -1, in agreement with the numerical results and the
analytical predictions. The data was log-binned, to reduce the
uneven statistical fluctuations common in heavy tailed
distributions, a procedure that does not alter the slope of the
tail. For the full curve, including the $\tau=1$ peak, see Fig. 3
in the Supplementary Information. {\bf (b)} Linear-log plot of the
$P(\tau)$ distribution for $p=0.00001$, corresponding to the
random choice limit of the model. The fact that the curve follows
a straight line on a linear-log plot indicates that $P(\tau)$
decays exponentially. }} \label{fig3}
\end{figure}

\end{document}